\documentclass[prd,aps,amsmath,amssymb,nofootinbib,twocolumn,preprintnumbers]
{revtex4}

\usepackage[utf8x]{inputenc}
\pdfoutput=1
\usepackage{amssymb,amsmath,mathrsfs,enumerate}
\usepackage{graphicx,rotate}
\usepackage{float}
\usepackage{tocloft}
\usepackage{subfig}
\usepackage{multirow}
\usepackage{slashed,xcolor}
\usepackage[margin=10pt,labelfont=bf]{caption}
\usepackage{soul}
\usepackage{setspace}
\usepackage[export]{adjustbox}

\makeatletter
\let\Hy@linktoc\Hy@linktoc@page
\makeatother

\usepackage{color}
\definecolor{ourcolor}{rgb}{0.7, 0.25, 0.05}
\usepackage{tikz,braket}

\long\def\rpl#1!!#2!!{\textcolor{red}{#1} \textcolor{blue}{#2}}

\definecolor{My_red}        {cmyk}{0.00,1.00,1.00,0.10}

\def\slash#1{\rlap/#1}
\let\tilde=\widetilde
\let\hat=\widehat
\let\bar=\overline

\def \order(#1){{\mathcal O} \left(#1 \right)}

\allowdisplaybreaks

\begin{document}

\preprint{HRI-RECAPP-2019-008}

\title{Impact of a colored vector resonance on the collider constraints for top-like top partner}

\author{Sayan Dasgupta$^{a,}$\footnote{sayandg05@gmail.com}}
\author{Santosh Kumar Rai$^{b,}$\footnote{skrai@hri.res.in}}
\author{Tirtha Sankar Ray$^{c,}$\footnote{tirthasankar.ray@gmail.com}}
\affiliation{$^{a}$Centre for Theoretical Studies, Indian Institute of Technology Kharagpur,
	Kharagpur 721302, India,\\
	$^b$Regional Centre for Accelerator-based Particle Physics, 
	Harish-Chandra Research Institute, HBNI, Chhatnag Road, Jhusi, Prayagraj 211019, India,\\
    $^c$Department of Physics, Indian Institute of Technology Kharagpur,
    Kharagpur 721302, India}

\begin{abstract}
In this work we reappraise  the  collider constraints from leptonic final states on the vectorlike  colored top partners taking into account the  impact of exotic colored vector resonances. These colored states are intrinsic to a broad class of models that employ a strongly interacting sector to drive electroweak symmetry breaking. We translate the recent results in the 
{\sl monolepton + jets} channel as reported by CMS with an integrated luminosity of 35.8 fb$^{-1}$, and {\sl dilepton + jets} and {\sl trilepton + jets} channels as reported by ATLAS with an integrated luminosity of 36.1 fb$^{-1}$ to constrain the parameter space of these class of models. We also comment on the impact and modification of the derived constraints  due to the expected fatness of the colored vector resonance, when accounted for beyond the narrow-width approximation by simulating the full one-particle irreducible resummed propagator.
\end{abstract}

\maketitle
\flushbottom

\section{Introduction}

The Large Hadron Collider (LHC) experiment is mandated to search for new physics beyond the Standard Model (SM) at the energy frontier. These discoveries are primarily expected to precipitate through unearthing of exotic states. In this hunt for exotics the colored vector gauge bosons and colored vectorlike fermions are low lying fruits. While they have large production cross section owing to their colored charges, they conveniently can be made  consistent with  electroweak observables measured at the $Z$-pole at the Large Electron-Positron (LEP) collider experiment \cite{Cacciapaglia:2010vn}. This may be contrasted with any extra chiral fermion generation which is heavily constrained by the electroweak observables. These states naturally arise in a class of well-motivated extensions of the SM-like extra-dimensional scenarios  \cite{Gherghetta:2006ha} and composite Higgs framework where the Higgs is identified with a psudo Nambu-Goldstone boson (pNGB) of the strong sector  \cite{Contino:2010rs}.  In situations where the colored vectorlike fermions participate in stabilizing the Higgs sector against quadratic sensitivity to the UV, they are usually labeled as top partners \cite{DeSimone:2012fs, Buchkremer:2013bha}. A huge cache of literature has built up regarding the phenomenology of the top partners \cite{Kim:2018mks, Kim:2019oyh, Moretti:2016gkr, Cacciapaglia:2018qep, Alhazmi:2018whk, Durieux:2018ekg, Cacciapaglia:2018lld, Yepes:2018dlw, Carvalho:2018jkq, Liu:2018hum, Barducci:2017xtw, Deandrea:2017rqp, Liu:2017sdg, Liu:2016jho, Matsedonskyi:2015dns, Barducci:2015vyf, Backovic:2015bca, Chala:2014mma, Basso:2014apa, Matsedonskyi:2014mna, Backovic:2014uma, Gripaios:2014pqa, Han:2014qia, Karabacak:2014nca, Andeen:2013zca, Azatov:2013hya, Banfi:2013yoa, Li:2013xba, Barcelo:2011wu, Barcelo:2011vk}.  In principle the  color triplet top partners can be in any representation of the weak gauge group but only certain combinations can mix with the SM top in the presence of the SM Higgs doublet.  In the context of stabilizing the Higgs sector the relevant representations are the so-called top-like multiplets that have at-least one state with quantum numbers identical to the  top that enables linear mixing between the top and the top partner. Focusing on the smaller representations, it is easy to see that the singlet and triplet top like multiplets will lead to stronger mixing with the SM top doublet, leading to stronger constraints from the electroweak observables  \cite{Cacciapaglia:2010vn}.  In this paper we will instead focus on the top-like top partner that is part of a doublet  and primarily mixes with the SM top right.

 We will consider the impact of any accompanying colored vector resonances on the constraints on the top partners from collider searches at LHC extending the study done in Refs.~\cite{Azatov:2015xqa, Araque:2015cna}.  While our phenomenological model remains agnostic to the specific UV realization, a large class of models including the extra-dimensional models and composite Higgs framework lead to scenarios that simultaneously have a top partner and heavy colored vector resonances \cite{Barnard:2013hka}. An interesting  facet of these models is the possibility of these exotic states being broad resonances.  Typically a state whose decay width is a sizable proportion $(> 20\%)$ of its mass is considered a broad resonance and the narrow width description starts to fail in maintaining gauge invariance. The large decay width can either be a consequence of large proliferation of the possible decay channels or a large nonperturbative coupling.  In this paper we will assume that the colored vector resonance, the  so-called gluon partner,  has a strong coupling with the top partner, inheriting  this from a strongly interacting sector they belong to.  This can be considered a pared down version of strongly interacting models of electroweak symmetry breaking, like the composite Higgs framework where the Higgs is a pNGB that couples to SM fermions through partial compositeness \cite{Contino:2010rs}.
  
We reappraise the present status of the  top-like top partners, that is, the vectorlike fermions having the same quantum numbers as the SM top,  in the light of the Run-2 results from LHC. In this context we recast the constraints on the parameter space of these scenarios from the searches for exotics in the leptonic final states at ATLAS and CMS. We systematically translate the relevant and most recent bounds from  CMS monolepton  study \cite{Sirunyan:2017pks} and ATLAS dilepton  and trilepton  \cite{Aaboud:2018saj} searches made in the context of vectorlike quarks.  We show that the exclusion limit on the top partners is moderately  altered due to the presence of the gluon partners. Additionally, the large width effect of the gluon partner is considerable and reconstructing the full one-particle irreducible (1PI) propagator for the \textit{fat} vector boson is quantitatively significant in most regions of the parameter space of  interest. We have compared the results obtained within the narrow width approximation to the ones obtained by simulating the full 1PI resummed propagator to demonstrate this.

The rest of the  paper is organized as follows. In Section \ref{sec:model} we introduce the phenomenological Lagrangian for the top partner and the gluon partner.  In Section \ref{beyond:nwa} we discuss the  impact of the large width of the vector resonance. In Section \ref{lhc:const} we  systematically translate the constraints on the parameter space of the model from LHC studies in leptonic final states before concluding.

\section{Model Lagrangian}\label{sec:model}

In this section we introduce the phenomenological Lagrangian involving the \textit{top-like} top partner  and  a colored vector  boson.   We extend the SM with a new  vectorlike colored fermion $\Psi \,(3,2,7/6)=\{X,U\}$  with mass $M$ and a  colored vector boson (the gluon partner) $\rho_{\mu}$, having mass $M_{\rho}$ and a large width $\Gamma_{\rho}$. A possible origin of such a spectrum in the context of a bottom-up composite Higgs framework is briefly sketched in Appendix~\ref{appn:chframework}. Concentrating on the SM third generation, the  new state $U$ will mix with the right-handed top that is assumed to be a member of a separate strongly interacting sector along with the exotic $X$ and $\rho_\mu$. The Lagrangian after electroweak symmetry breaking can be parameterized as shown in Equation~\ref{master:lgn} \cite{Cacciapaglia:2011fx},
where, $q_L=\{\tilde{t_L},b_L\}^T$ and two singlets $\tilde{t_R}$ and $b_R,$ are the usual third generation SM  quarks in the gauge basis. The mixing between $U_L$ and $\tilde{t_R}$ is assumed to originate from an underlying Yukawa coupling in the strongly interacting sector. The covariant derivatives beside containing the usual SM gauge 
interactions of the colored fermions, include the coupling to the massive colored vector boson $\rho_\mu,$ given by 
\begin{equation}
\slashed{D} \supset -ig_{i}\slashed{\rho} 
\label{eq:covder}
\end{equation}
where $g_{i}= g_\ast,$ for the strong sector resonances viz.  $\tilde{t_R}$ and  $U,$ while $g_{i}$=-$g_s^2/g_{\ast}$ for the  elementary states. 
\begin{widetext}
	\begin{equation}
	\mathcal{L}_{eff} \supset i\bar{\Psi}\slashed{D}\Psi+i\bar{q_L}\slashed{D}
	q_L+i\bar{\tilde{t_R}}\slashed{D}
	\tilde{t_R}+\bar{b_R}\slashed{D}
	b_R+\frac{1}{2}M_{\rho}^2\rho^{\mu}\rho{\mu} - \big[ \tilde{m_t}\bar{\tilde{t_L}}\tilde{t_R}+m_{mix}\bar{U_L}\tilde{t_R}+M\bar{U_L}U_R + h.c.\big],  
	\label{master:lgn}
	\end{equation}
\end{widetext}
 The latter is a special choice adopted assuming a 5d gauge-Higgs UV completion of these 
models \cite{Barnard:2013hka}.
In the mass basis the mass terms can be written as,
\begin{equation}
\mathcal{L}_{mass}=-m_t\bar{t_L}t_R-m_{t^{\prime}}\bar{t_L^{\prime}}t_R^{\prime}+h.c.
\end{equation}
where $t$ represents the SM top and $t^\prime$ is the heavier top-like top partner. The corresponding rotation matrices  can be schematically written as
\begin{equation}
\begin{pmatrix}
t_L \\ t_L^{\prime} \\ t_R \\ t_R^{\prime}
\end{pmatrix}
=
\begin{pmatrix}
U_{\theta_L} & 0 \\
0 & U_{-\theta_R} 
\end{pmatrix}
\begin{pmatrix}
\tilde{t}_L \\ U_L \\ \tilde{t}_R \\ U_R  
\end{pmatrix}
\label{eq:mixingmatrix}
\end{equation}
where $U_{\theta_i}$ are the 2D rotation matrices. The parameters and the mixing angles are correlated as
\begin{equation}
\begin{split}
&\sin\theta_R= \frac{M m_{mix}}{\sqrt{(M^2-m_t^2)^2+M^2 m_{mix}^2}}\hspace{0.1in}=\frac{M}{m_t}\sin\theta_L  \\\\
&\mbox{where,}~~ \hspace{0.1in}M^2=\frac{m_{t^{\prime}}^2+\sin^2\theta_Rm_t^2(m_{t^{\prime}}^2-m_t^2)}{1+\sin^2\theta_R(m_{t^{\prime}}^2-m_t^2)}
\end{split}
\label{eq:paramrelations}
\end{equation}

The couplings of $\rho_{\mu}$ (in the mass basis) with the top and top partner can be read out from Equations~\ref{master:lgn}-\ref{eq:paramrelations},
\begin{equation}
\begin{split}
\mathcal{L}_{\rho tt^{\prime}} = & (g_{\ast}\sin^2\theta_L-\frac{g_s^2}{g_{\ast}}\cos^2\theta_L)\bar{t}_L\slash{\rho}t_L \\&+ (g_{\ast}\cos^2\theta_L-\frac{g_s^2}{g_{\ast}}\sin^2\theta_L)\bar{t}_L^{\prime}\slash{\rho}t_L^{\prime} \\
& - \frac{\cos\theta_L\sin\theta_L}{g_{\ast}}(g_s^2+g_{\ast}^2)(\bar{t}_L^{\prime} \slash{\rho} t_L+\bar{t}_L \slash{\rho} t_L^{\prime}) \\&+ g_{\ast}(\bar{t}_R \slash{\rho} t_R+\bar{t}_R^{\prime} \slash{\rho} t_R^{\prime}) \,\, .
\end{split}
\label{eq:rhocoup}
\end{equation}
This effective framework has  $m_{t^{\prime}}$, $\theta_R$, $g_{\star}$ and $M_{\rho}$ as the free parameters of the model. However to keep the discussion tractable we will consider a benchmark scenario where  $g_{\star} / g_s$ will be set 
at $6$ \cite{Carena:2007tn} which is in good agreement with large-N calculations in the strongly interacting  holographic dual  theory of a pNGB composite Higgs model \cite{Barnard:2013hka}.

At the LHC the top partner $t^{\prime}$ is pair produced through  the  gluon or through the massive gluon partner ($\rho_{\mu}$). Once they are produced they will dominantly  decay to SM states through the channels: $Ht, Zt$ and $Wb.$ The branching ratios of $t^{\prime}$ at $m_{t^{\prime}} \sim 1$ TeV in the main decay channels are given by: BR($t^{\prime}\rightarrow H \, t)=0.56$, BR($t^{\prime}\rightarrow Z \, t)=0.42$ and BR($t^{\prime}\rightarrow W \, b)=0.02$  \cite{Cacciapaglia:2011fx}.  The reduced branching ratio to the $Wb$ is a consequence of the exotic state $U$ primarily mixing with the $SU(2)_L$ singlet state $\tilde{t_R}$.
Here we assume that there are no significant exotic decays of the top partners \cite{Bizot:2018tds}. Here and for the rest of this paper we set  $\sin\theta_R=0.1$ which is a conservative choice keeping the  framework relatively insulated from the electroweak precision constraints \cite{Cacciapaglia:2011fx}.  The choice of the strong sector coupling $g_\ast$ and 
the mixing angle $\sin\theta_R$ forms a benchmark scenario that will be utilized in all the phenomenological studies that follow.

\section{Beyond the Breit-Wigner} \label{beyond:nwa}
\begin{figure*}[t]
	\centering
	\subfloat[\label{fig:cross1}]{\includegraphics[scale=0.5]{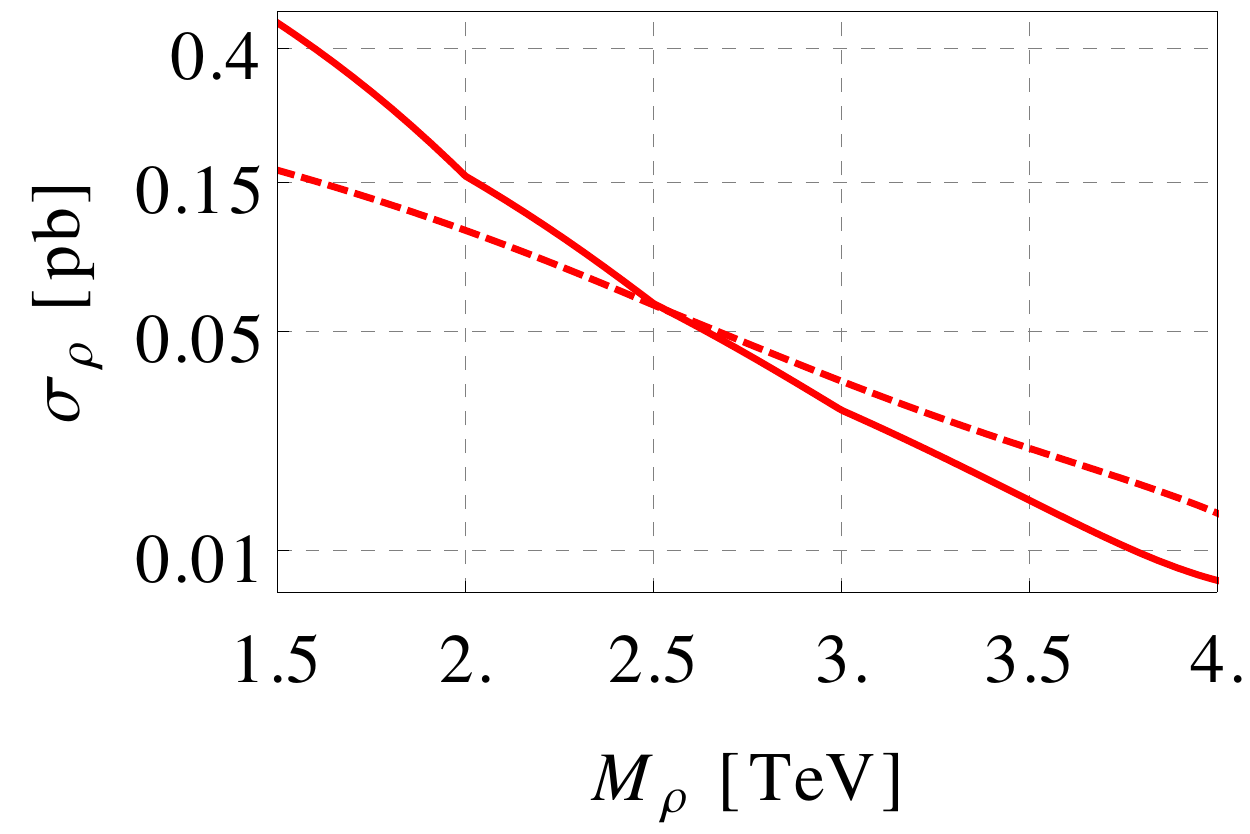}}
	\hspace{0.4cm}
	\subfloat[\label{fig:cross2}]{
		\includegraphics[scale=0.53]{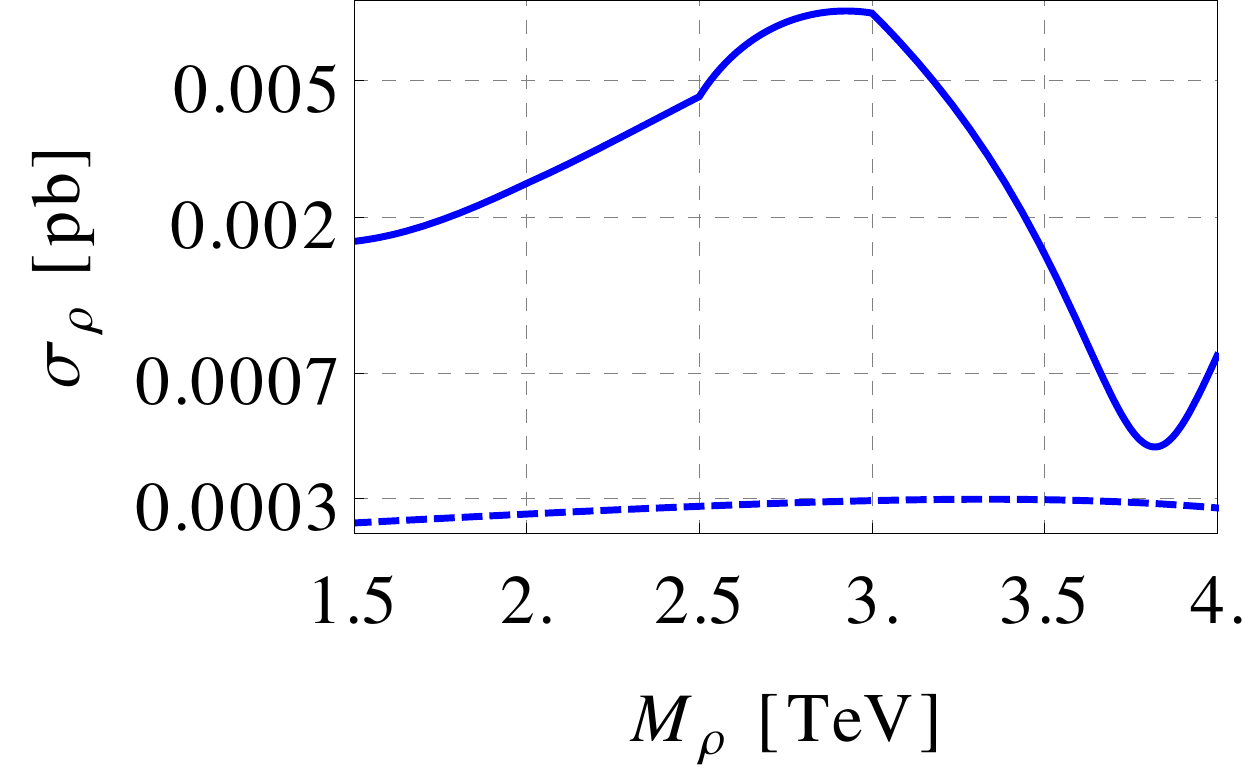}}
	\caption{$\sigma_{\rho}$ obtained from Breit-Wigner approximation (solid) and by calculating the full 1PI resummed propagator (dashed) for (a) $m_{t^{\prime}}$=0.5 TeV and (b) $m_{t^{\prime}}$=1.5 TeV.}
	\label{fig:cross}
\end{figure*}

\begin{figure}[t]
	\centering
	\includegraphics[scale=0.6]{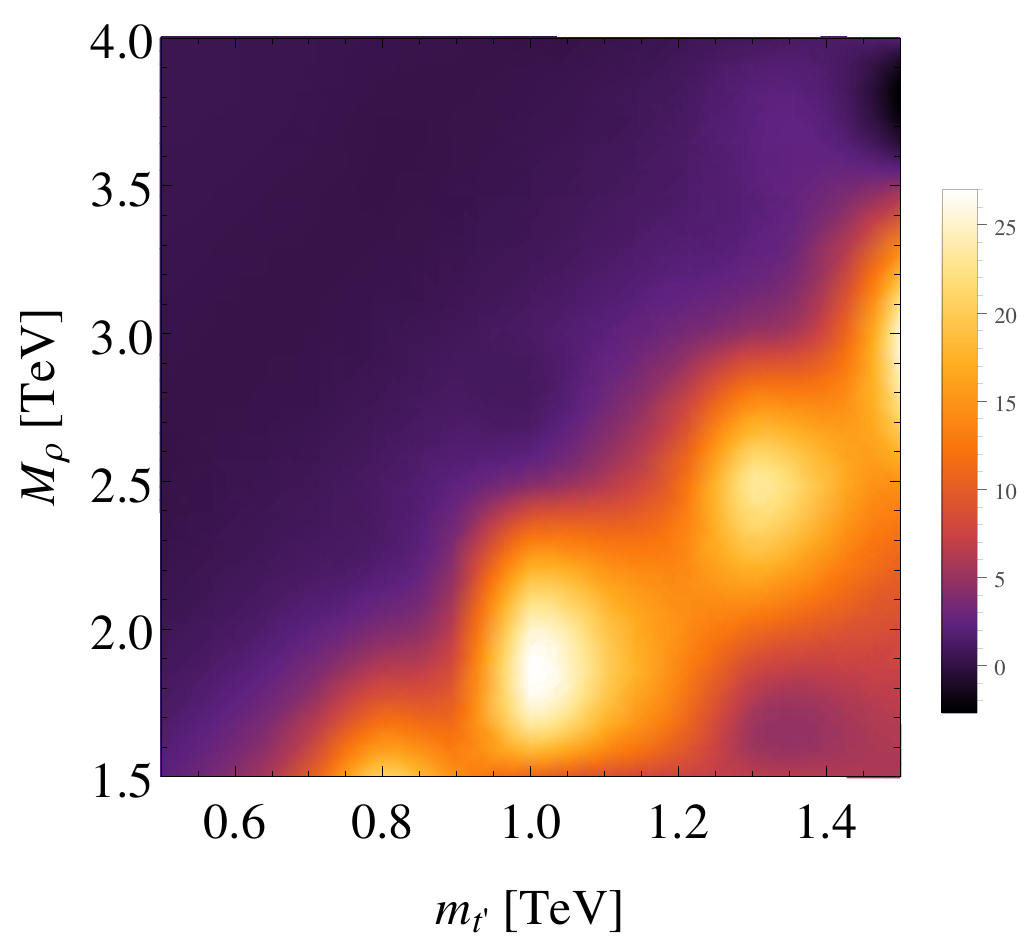}
	\caption{Density plot depicting the impact of fatness correction ($\mathcal{F}$ as defined in Equation~\ref{eq:crosserrorfrac}) in the parameter space of interest.}
	\label{fig:cross-error}
\end{figure}

In the parameter space of interest the total decay width of $\rho_{\mu}~(\Gamma_{\rho})$ consistently remains above $20\%$ of its mass ($M_{\rho}$) for the choice of $g_\ast,$ where the decay to a pair of top partner is kinematically possible. In this region, the Breit-Wigner (BW) approximation may not be a good approximation and starts to fail. The usual gauge invariant approach  to handle broad resonances is the complex mass scheme \cite{Denner:1999gp}; however, for  massive vector resonances there is no gauge invariance issue with  large decay width and basically maps into the usual narrow width results with the appropriate  enlarged value of the decay width in the usual BW propagator. However careful analysis should include the impact of the large width by utilizing the full 1PI  propagator in computations of the cross section. To systematically handle this we recalculate the top partner production cross section using the full 1PI resummed propagator for the 
$\rho_{\mu}$ \cite{Azatov:2015xqa}. 
The pure $\rho_\mu$ contribution to the $t^{\prime}\bar{t^{\prime}}$ production cross section is
\begin{equation}
\sigma_{\rho}^{\rm fat}=2\int_{0}^{1}d\tau \hat{\sigma}(S^{\rm had}\tau)_{\rm fat}\int_{\tau}^{1}\frac{dx}{x}\sum_{q} f_q(x)f_{\bar{q}}(\frac{\tau}{x})
\end{equation}  
where $S^{\rm had}$ is the hadronic center of momentum  energy, $\hat{\sigma}$ is the partonic cross section and the functions $f_{q/\bar{q}}$ are parton density functions.
For the pair production of $t^{\prime}$ through an $s$-$channel$ $\rho_{\mu}$ exchange at LHC (including the full 1PI resummed propagator for the $\rho_{\mu}$), the partonic cross section is
\begin{equation}
\begin{split}
\hat{\sigma}(\hat{s})^{\rm fat}=&\frac{g_{\rm prod}^2 g_{\rm dec}^2}{27\pi \hat{s}}\sum_{\chi}\frac{\sqrt{\hat{s}(\hat{s}-4m_{\chi}^2)}}{(\hat{s}-M_{\rho}^2)^2+(Im[M^2(\hat{s})])^2}\\&\times (\hat{s}+2m_{\chi}^2),
\hspace{0.5cm}
\chi=t,t^{\prime}
\end{split}
\label{eq:fatcross}
\end{equation}
The imaginary part  of $M^2(\hat{s})$ in the above expression represents the contribution from one loop corrections to the  $\rho_{\mu}$ propagator. Since in the model $t_R$ is assumed to be  a state in the strong sector, both  t and $t^{\prime}$ will  contribute in the loop  and the relevant expression is given by
\begin{equation}
Im[M^2(\hat{s})]=-\frac{g_{\rm dec}^2}{12\pi \sqrt{\hat{s}}}\theta(\sqrt{\hat{s}}-2m_{\chi})\sqrt{\hat{s}-4m_{\chi}^2}(\hat{s}-m_{\chi}^2)
\label{eq:fatim}
\end{equation} 
The difference in $\sigma_{\rho}$  ($pp\rightarrow \rho \rightarrow t^{\prime}\tilde{t^{\prime}}$ cross section) obtained from Breit-Wigner approximation ($\sigma_{\rho}^{narrow}$) and by calculating the full 1PI resummed propagator is visible from the plots displayed in Figure~\ref{fig:cross}.
In this work we have implemented the full 1PI resummed propagator for the fat resonances as defined in Equations \ref{eq:fatcross} and \ref{eq:fatim} into {\tt MadGraph5} to simulate the impact of arbitrary virtualness of such resonances in collider studies. This is in contrast with previous studies \cite{Azatov:2015xqa} where the impact of the 1PI loop was considered by modifying the cross section. This approach neglects the impact of interference and misses the impact of fatness on the final state kinematic shapes which can potentially modify the efficiencies. The cross section modulation factor $\mathcal{F}$ is defined as
\begin{equation}
\mathcal{F}=\frac{|\sigma^{\rm fat}_{\rho}-\sigma^{\rm narrow}_{\rho}|}{\sigma^{\rm fat}_{\rho}}
\label{eq:crosserrorfrac}
\end{equation}
This has been plotted in the parameter space of interest in Figure~\ref{fig:cross-error}. From the plot one can easily read off that the departure of the two cross sections peaks around the resonance ($M_{\rho}=2m_{t^{\prime}}$) and in most of the parameter space of interest it stays low ($<$10\%).

\section{LHC Constraints} \label{lhc:const}
The effective framework  described in Section~\ref{sec:model} has been simulated  by writing a model file in {\tt Feynrules 2.0} \cite{Alloul:2013bka} and a {\tt UFO} file was generated. The values of the free parameters used in the analysis have been summarized in Table~\ref{tab:parameters}.
\begin{table}
	\centering
	\begin{tabular}{ |c|c|c|c| } 
		\hline
		$m_{t^{\prime}}$ & $M_{\rho}$ & $g_{\ast}$ & $\sin\theta_R$ \\ 
		\hline
		0.5-1.5 TeV & 1.5-4.0 TeV & 7.317 & 0.1 \\ 
		\hline
	\end{tabular}
\caption{Values of free parameters of the model used in the analysis.}
\label{tab:parameters}
\end{table}
This was imported in {\tt MadGraph5} \cite{Alwall:2014hca} and  pair production events of the top partner $t^{\prime}$ were generated.
Events were parton-showered using {\tt Pythia8}\,\cite{Sjostrand:2014zea}, jet-clustered using {\tt FastJet}\,\cite{Cacciari:2011ma} and passed through detector simulation using {\tt Delphes-3}\,\cite{deFavereau:2013fsa}. Note that {\it object reconstructions} have been done using the default cards available in {\tt Delphes}, modified where necessary. Four different LHC searches were used to constraint the model {\it viz.}, monolepton + jets \cite{Sirunyan:2017pks}, dilepton + jets + large-R jets (1 large jet and $\geq$ 2 large jets) \cite{Aaboud:2018saj}, and trilepton + jets \cite{Aaboud:2018saj}. The recast for each was written in {\tt MadAnalysis5}\,\cite{Conte:2012fm} and the efficiencies were obtained.  To obtain the $95\%$ exclusion we used the following generic template
\begin{equation}
\sigma(m_{t^{\prime}},\sin\theta_R,M_{\rho})\times \epsilon\times \mathcal{L} \leq \rm N_{\rm signal}
\label{eq:Nsim}
\end{equation}
where $\epsilon$ is the efficiency obtained by applying the cuts on the generated signal events, $\mathcal{L}$ is the integrated luminosity at which the LHC analyses were reported and $\rm N_{\rm signal}$ is the $95\%$ exclusion bound on the total number of simulated events presented in the analysis and $\sigma$ is pair production cross section of $t^{\prime}$ obtained from {\tt MadGraph5} multiplied by the corresponding $K\sim 1.4$ obtained from {\tt Top++2.0}\,\cite{Czakon:2011xx} at next-to-leading order. These are conservative choices which are in consonance with the estimates  for the relevant $K$-factor quoted in Ref.~\cite{Cacciari:2008zb} and \cite{Zhu:2012um} for SM $t\bar{t}$ and Kaluza-Klein (KK) gluon mediated $t\bar{t}$ production at LHC. Since we have included the contribution from fatness of $\rho$ by modifying the propagator inside {\tt MadGraph5}, we did not have to neglect the contribution to cross section from the interference of the production processes of $t^{\prime}\bar{t^{\prime}}$ through $\rho$ and QCD which in some regions of the parameter space of interest can be quite large as visible in Figure~\ref{fig:exclusion_total_interference}. Further, we do not make any simplifying assumption regarding the branching ratios  of $t^{\prime}$ and we keep all decay channels as mentioned in Section~\ref{sec:model}.  

Additional bounds on $\rho_{\mu}$ from direct searches for KK-gluon, through $t\bar{t}$-production in multileptonic and hadronic channels by CMS \cite{Sirunyan:2018ryr} have been translated to the parameter space of the model. For this we simply translate the 
bound on the cross section without recasting the experimental search. However we 
have taken care of the mass dependent decay branching ratios of the gluon partner. 

In the rest of this section we systematically study the constraints on the benchmark model parameter space from various channels having leptonic final states. It is worth pointing out here that  the total decay width of $\rho$ is a fast growing quantity and quickly rises close to the mass value of the field itself. In such a scenario, providing a particle interpretation with on-shell production of such a state may not be feasible. Therefore,  to study the effect of fatness and without a prejudice on how fat it becomes, we have kept the width of $\rho$ far greater than 20\% while avoiding a particlelike interpretation for this state and simply consider its role as a quantum field.

\subsection{Monolepton+jets}
\begin{figure*}[t]
	\centering
	\subfloat[\label{fig:process_feyn_mono}]{\vspace{-2cm}
		\adjincludegraphics[trim={5cm {.78\width} 5cm 0},clip ,scale=0.6]{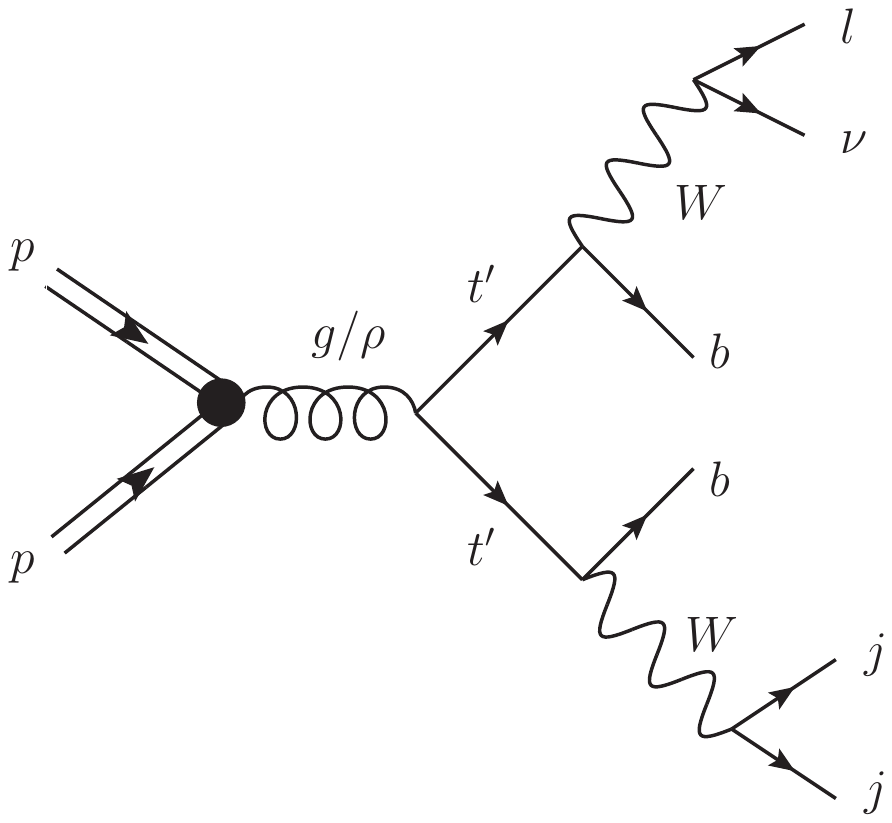}}
	\hspace{1cm}
	\subfloat[\label{fig:exclusion_mono}]{
	\includegraphics[scale=0.5]{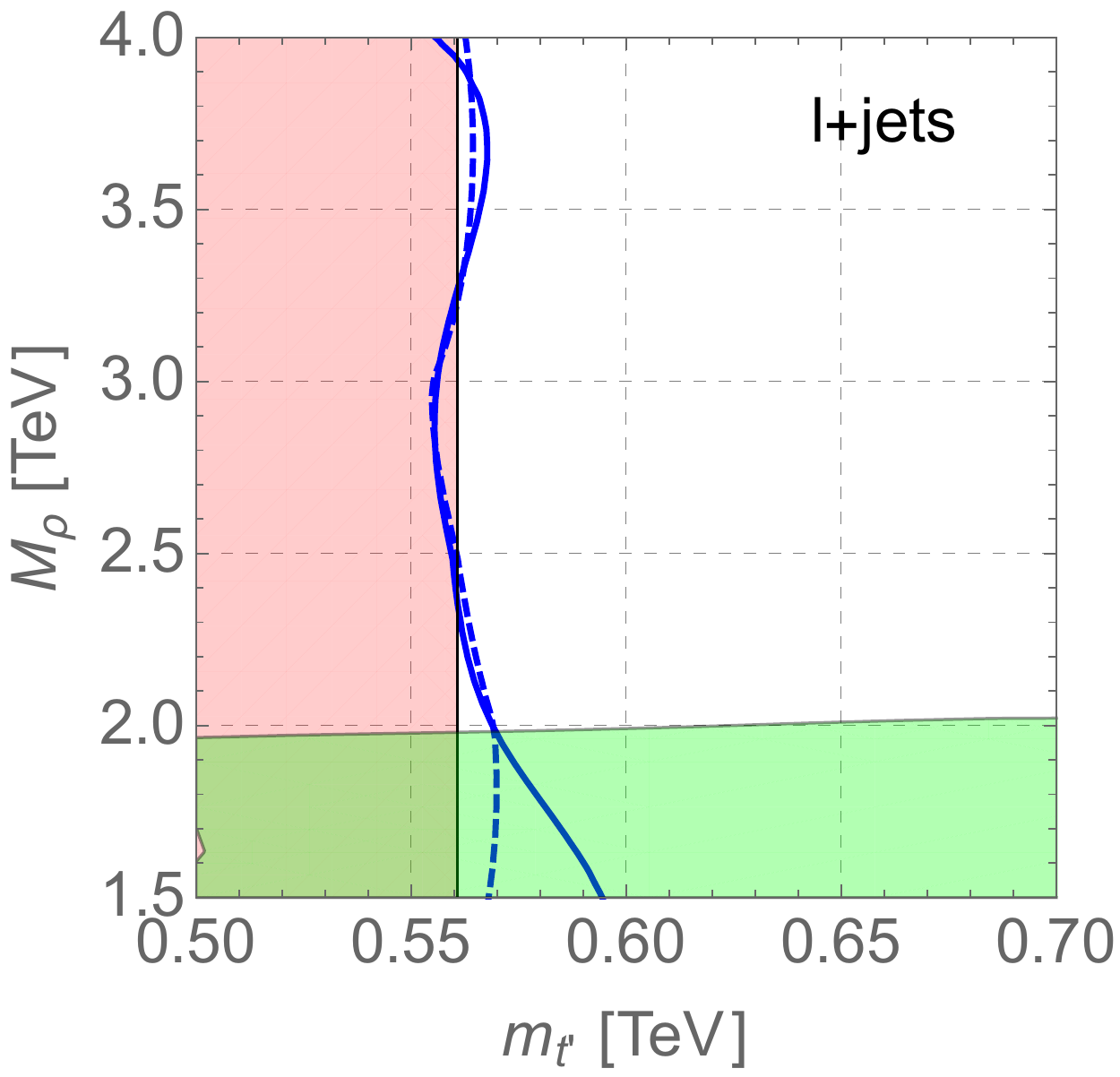}}
	\caption{(a) Feynman diagram and (b) $95\%$ C.L. exclusion contours from the monolepton channel on the ($m_t^{\prime},M_{\rho}$) plane. In (b) solid blue line is the plot for NWA, dashed blue line represents the plot for fat-width correction, the red excluded region corresponds to the QCD production of the top partner, and the green region is the $95\%$ C.L. exclusion region from KK-gluon search. The regions to the left of the contours are excluded.}
\end{figure*}
In this section we summarize the constraints on the parameter space of our benchmark model based on the CMS l+jets study  with an integrated luminosity of 35.8 fb$^{-1}$ for the muon channel and with an integrated luminosity of 35.6 fb$^{-1}$ for the electron channel \cite{Sirunyan:2017pks}. The CMS analysis considers the decay of the top partner vectorlike quark (VLQ) in the $b \, W$ mode only. Unlike signal production for the other analyses presented in this paper, for this case we force the top partner to decay to W and b in the event generation level. We thus neglect the contribution to the final signal topology from other decay modes which are supposed to be largely suppressed. The similar simulated signal process in our model is shown in Figure~\ref{fig:process_feyn_mono}. The search focuses on two W-bosons and two b-jets one of which decays leptonically giving rise to a single lepton and the other decays hadronically. The dominant background (SM) subprocesses which contribute to the same signal channel are $t\bar{t}$, $W$+jets and single top production. Preselection requires all events to have missing $p_T$ greater than 30 GeV. Charged leptons (electron or muon) were required to have a minimum $p_T$ of 55 GeV and a maximum absolute pseudorapidity ($|\eta|$) of 2.4. Jets reconstructed using anti-$k_T$ algorithm and having a minimum $p_T$ of 30 GeV and a maximum $|\eta|$ of 2.4 were collected. Out of these the jets too close to the charged lepton ($\Delta R(jet,l) < 0.4$) were removed. At least four jets were required after this removal. The first and second highest $p_T$ jets were required to have minimum $p_T$ of 100 and 70 GeV respectively. Note that the collaboration separately presented their results in the electron and muon channels and provided a combined bound on the top partner mass. We separately simulated the efficiencies in the electron and the muon channels and obtained the combined events for each point in the simulated parameter space. This was compared with the 95\% uncertainty in the simulated background events presented in  \cite{Sirunyan:2017pks} obtained by adding in quadrature the individual uncertainties in the electron and muon channels. The cuts applied to mimic the signal region of \cite{Sirunyan:2017pks} are described below.
\begin{itemize}
	\item All events were required to have exactly one charged lepton (electron or muon) and four jets which have been clustered using anti-$k_T$ algorithm with a radius parameter 0.4. Of the four jets, two were required to be b-jets arising from $t^{\prime}$ decays. We named the one accompanying the $W$ boson which decays leptonically $b_l$ and the one accompanying the $W$ boson which decays hadronically $b_h$. The remaining two jets were named $j_1$ and $j_2$ according to their $p_T$.
	\item The $b_h$, $b_l$, $j_1$ and $j_2$ jets were required to have $p_T$ greater than 200, 100, 100 and 30 GeV respectively.
	\item A variable $S_T$ was defined as the scalar sum of missing $p_T$, $p_T$ of the signal lepton and $p_T$ of the four jets. All signal events were required to have $S_T$ greater than 1 TeV.
	\item A similar variable $S_L$ was defined as the scalar sum of the reconstructed longitudinal component of the neutrino momenta and the longitudinal component of the signal lepton and jets momenta. All events were required to have $S_L/S_T$ less than 1.5.
	\item The invariant mass of $j_1$ and $j_2$ were required to be in the range 60-100 GeV to ensure that they decayed from a $W$ boson.
	\item The invariant mass of the signal lepton, neutrino and $b_l$, and $j_1$, $j_2$ and $b_h$ were required to be matched with a 60\% accuracy to ensure that these two sets of objects originated from decaying $t^{\prime}$.
\end{itemize}
To validate the recast code of this search written in {\tt MadAnalysis5} we have generated SM $t\bar{t}+jets$ and $t\bar{t}+V$ (where V is either W or Z) events and matched the event numbers with those given in Table 1 of Ref.~\cite{Sirunyan:2017pks} with an average accuracy of 20\%. Also signal efficiencies for three values of $m_{t^{\prime}}$ were matched with those reported in Table 2 of ~\cite{Sirunyan:2017pks}. The $95\%$ C.L. exclusion contour from this analysis is shown in Figure~\ref{fig:exclusion_mono}. As can be seen from the plot, the constraints on $m_{t^{\prime}}$ become more severe than the QCD limit (shaded red) for $M_{\rho}< 2.0$ TeV, most of which is excluded from the direct limit on $M_{\rho}$ from the KK-gluon search (shaded green). Significant improvement from the QCD limit which is allowed can be seen around $M_{\rho}$=1.5 TeV. The effect of fatness reduces the impact of $\rho_{\mu}$. Due to low branching ratio of $t^{\prime}>bW$ in our model (falls off from 0.23 at $m_{t^{\prime}}$=0.5 TeV to 0.01 at $m_{t^{\prime}}$=1.5 TeV.), this search provides highly subdominant bounds compared to the other searches that we describe below. Note that the direct bound on $M_{\rho}$ from the kk-gluon search  saturates to $\sim 2.0$ TeV for smaller values $m_{t^{\prime}}$ in the region of interest.

\begin{figure*}[t]
	\centering
	\subfloat[\label{fig:process_feyn_di}]{\vspace{-2cm}
		\adjincludegraphics[trim={5cm {.9\width} 6.5cm 0},clip ,scale=0.7]{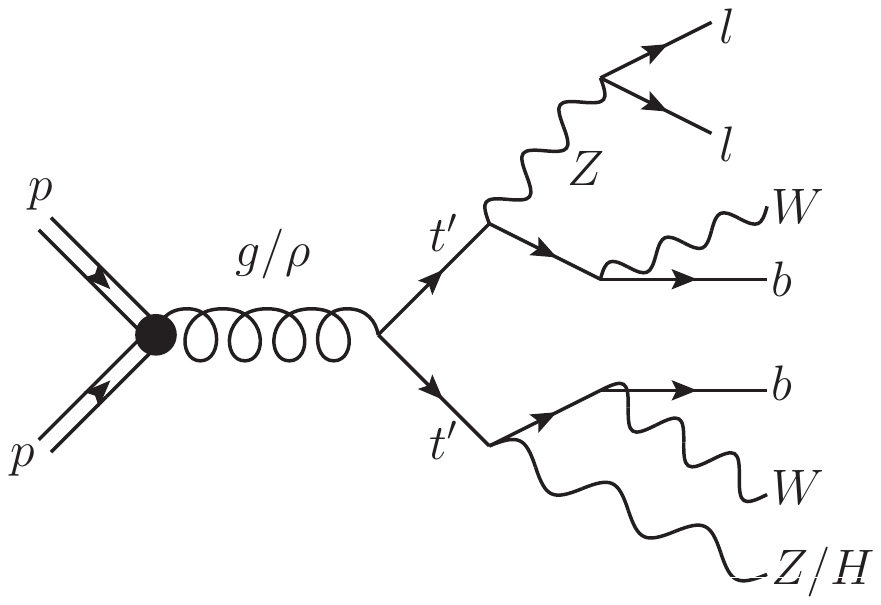}}
	\hspace{1cm}
	\subfloat[\label{fig:exclusion_di}]{
		\includegraphics[scale=0.5]{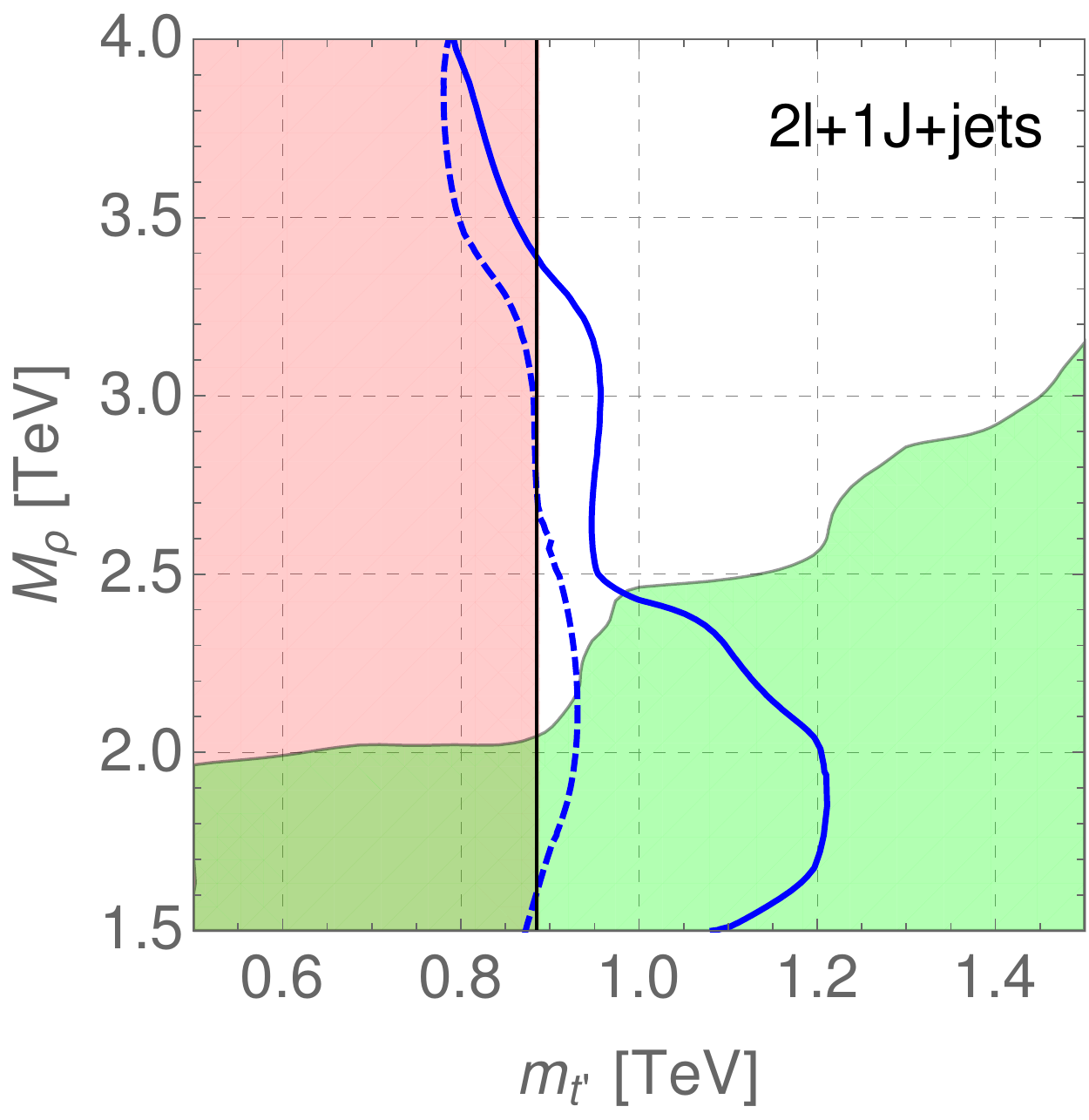}}
	\caption{(a) Feynman diagram and (b) $95\%$ C.L. exclusion contours from the dilepton channel on the ($m_t^{\prime},M_{\rho}$) plane. In (b) solid blue line is the plot for NWA, dashed blue line represents the plot for fat-width correction, the red excluded region corresponds to the QCD production of the top partner, and the green region is the $95\%$ C.L. exclusion region from KK-gluon search. The regions to the left of the contours are excluded.}
\end{figure*}

\subsection{Multilepton+jets}

In this section we summarize the constraints on the parameter space of our benchmark model from ATLAS 2l+jets and 3l+jets studies with an integrated luminosity of 36.1 fb$^{-1}$ \cite{Aaboud:2018saj} at $\sqrt{s}=13$ TeV . 
The ATLAS Collaboration has looked at the production of vectorlike quarks through the standard strongly interacting production channel and has put limits on their mass by looking at leptonic final states characterized by the presence of a reconstructed high $p_T$ $Z$ boson along with $b$-tagged jets.  This closely resembles the final states desired in our study.
Although the ATLAS study considers both single and pair production of the VLQs and construct nearly five different combinations of the signal region (SR) for their study of third generation partners of VLQ, we shall only focus on the signal regions which characterize our top-like partner pair production. Thus we choose only those relevant parts of the analysis which are likely to produce the maximum sensitivity to our signal analysis and reinterpret the results typical to the given choice. In all three analyses mentioned below charged leptons were required to have a minimum $p_T$ of 28 GeV. Electrons and muons were required to have maximum $|\eta|$ of  2.47 and 2.5 respectively. Small-R jets were required to have maximum $|\eta|$ of 4.5. Out of these the ones with $|\eta|$ less than 2.5 were required to have $p_T$ greater than 25 GeV and the ones with $|\eta|$ greater than 2.5 were required to have $p_T$ greater than 35 GeV. Large-R jets needed for the dilepton analyses were required to have a maximum $|\eta|$ of 2.0 and minimum mass and $p_T$ of 50 and 200 GeV respectively. These were constructed by combining the four momenta of small-R jets which were within a radius 1.0 having invariant mass greater than 50 GeV and combined $p_T$ greater than 200 GeV.
\begin{figure}[t]
	\centering
		\includegraphics[scale=0.5]{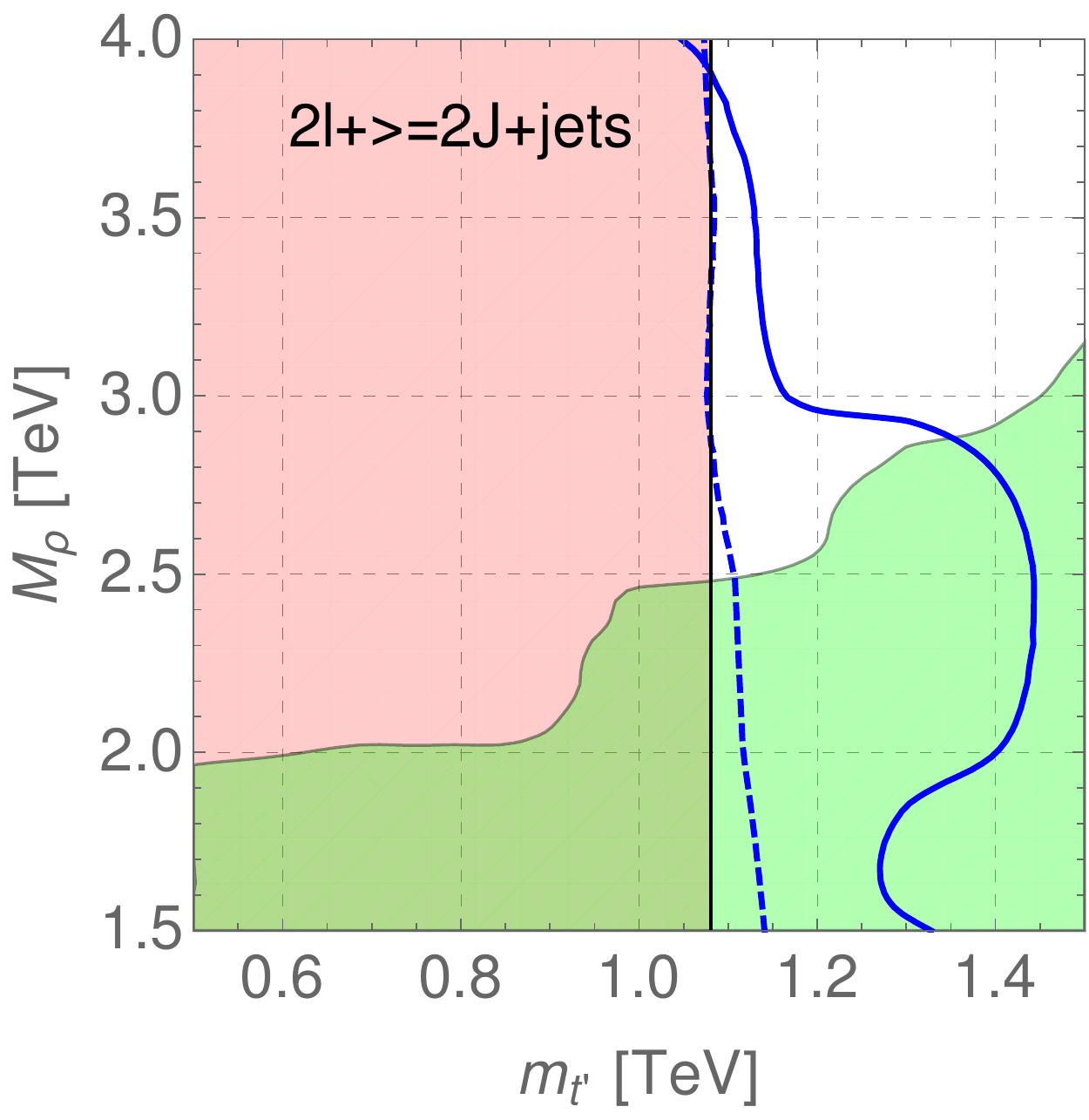}
	\caption{$95\%$ C.L. exclusion contours from the dilepton + $\geq$2J channel on the ($m_t^{\prime},M_{\rho}$) plane. Solid blue line is the plot for NWA, dashed blue line represents the plot for fat-width correction, the red excluded region corresponds to the QCD production of the top partner, and the green region is the $95\%$ C.L. exclusion region from KK-gluon search. The regions to the left of the contours are excluded.}
	\label{fig:exclusion_di_2J}
\end{figure}
\vspace{-0.5cm}
\subsubsection{Dilepton+1J+jets}
The simulated process for the signal in our model is depicted in Figure~\ref{fig:process_feyn_di}. The dominant SM background subprocesses which contribute to the 2l+1J+jets channel are $Z$+jets, $t\bar{t}$ and single top production. 

Note that both the dilepton and trilepton signals have a few preselection conditions common for the final states in the signal regions. Preselection requires that the final states have a $Z$ boson candidate which decays to a same flavor opposite sign lepton pair, such that the invariant mass of the dilepton must always be less than 400 GeV. Thus the final state event must have at least two charged leptons of same flavor and events with $M_{l^+l^-} > 400$ GeV are rejected. Note that this dilepton study by ATLAS divides the SR into several different categories where the number of large R-jets play a crucial role. The two signal regions (0-large-R jet SR and 1-large-R jet SR)\footnote{Small-R jets are reconstructed using {\tt Fastjet} with the anti-kt algorithm with the  radius parameter 0.4. Large-R jets were constructed by combining small-R jets within a radius 1.0.} described in Table 3 of Ref.~\cite{Aaboud:2018saj} in principle characterize the signal in our case and we choose the latter (which gives a better signal-to-background ratio) for our dilepton+1J analysis since the signal topology explored here is expected 
 to have one large-R jet as evident from Figure~\ref{fig:process_feyn_di}. The other signal region provides a subdominant contribution to the constraints on our parameter space and hence we choose to neglect it. Following the above-mentioned 
 signal region, we list the relevant kinematic selections on the final state events for our analysis: 
 \begin{itemize}
	\item All events were required to have exactly two oppositely charged same flavored leptons (electrons or muons) with $p_T$ greater than 28 GeV. Out of these the lepton pair decaying from the $Z$ boson was identified as the one with invariant mass closest to the $Z$ boson mass.
	\item The invariant mass of the same flavor charged lepton pair was required to 
	be within 10 GeV of the $Z$ boson mass.
	\item The transverse momentum of the charged lepton pair ($p_{T_{l^+l^-}}$) was required to be greater than 250 GeV. This is to exploit the 	high $p_T$ feature of the $Z$ boson which comes from the decay of the heavy top partner.
	\item All events were required to have at least two small-R jets which were clustered using anti-$k_T$ algorithm with a radius parameter 0.4 and had $p_T$ greater than 25 GeV. At least two of them were required to be $b$-tagged which help in suppressing the SM background coming from $Z$+jets. 
	\item All events were required to have exactly one large-R jet.
	\item A variable $H_T$ was defined as the scalar sum of the $p_T$ of all small-R jets and was required to be greater than 800 GeV.
\end{itemize}
To validate the recast code of this search written in {\tt MadAnalysis5} we have generated SM $Z$+jets and $t\bar{t}$ events and matched the cross section times efficiency times integrated luminosity with those given in Table 9 (1 large-R jet SR column) of Ref.~\cite{Aaboud:2018saj} within 10\% accuracy. The $95\%$ C.L. exclusion contour from this analysis is shown in Figure~\ref{fig:exclusion_di}. As can be seen from the plot, the constraints on $m_{t^{\prime}}$ become more severe than the QCD limit (shaded red) for $M_{\rho}<2.5$ TeV, most of which is excluded from the direct limit on $M_{\rho}$ from the KK-gluon search (shaded green). Significant improvement from the QCD limit which is allowed can be seen around $M_{\rho}$=2 TeV. The fatness of $\rho$ reduces its impact on the bound on $m_{t^{\prime}}$.
\begin{figure*}[t]
	\centering
	\subfloat[\label{fig:process_feyn_tri}]{\vspace{-2cm}
		\adjincludegraphics[trim={5cm {.8\width} 4.5cm 0},clip ,scale=0.6]{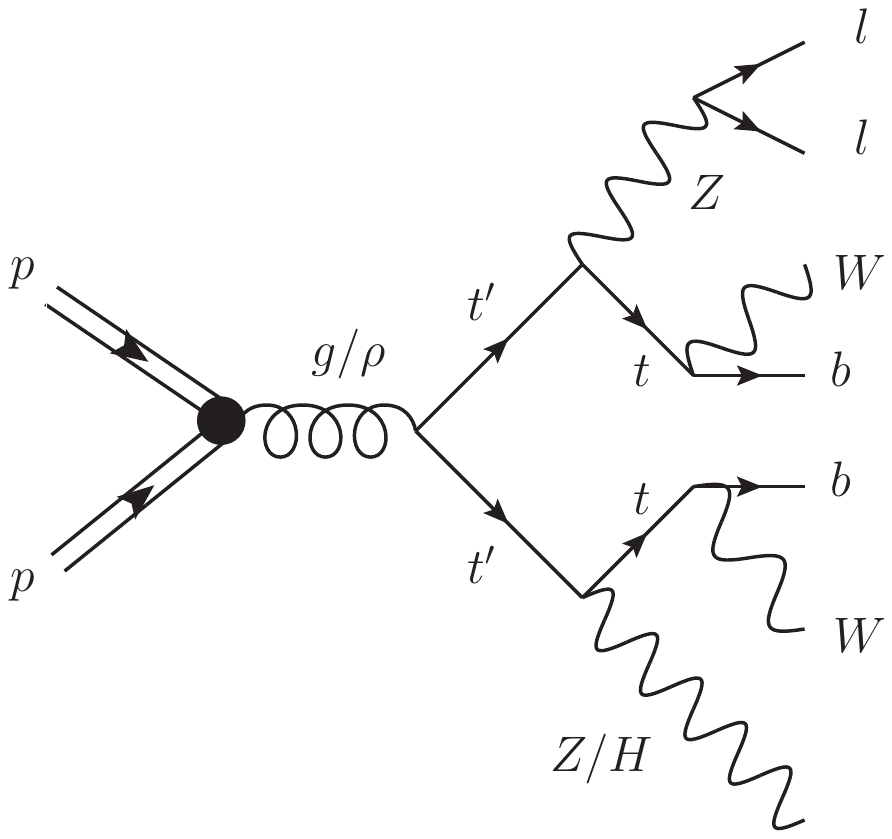}}
	\hspace{1cm}
	\subfloat[\label{fig:exclusion_tri}]{
		\includegraphics[scale=0.5]{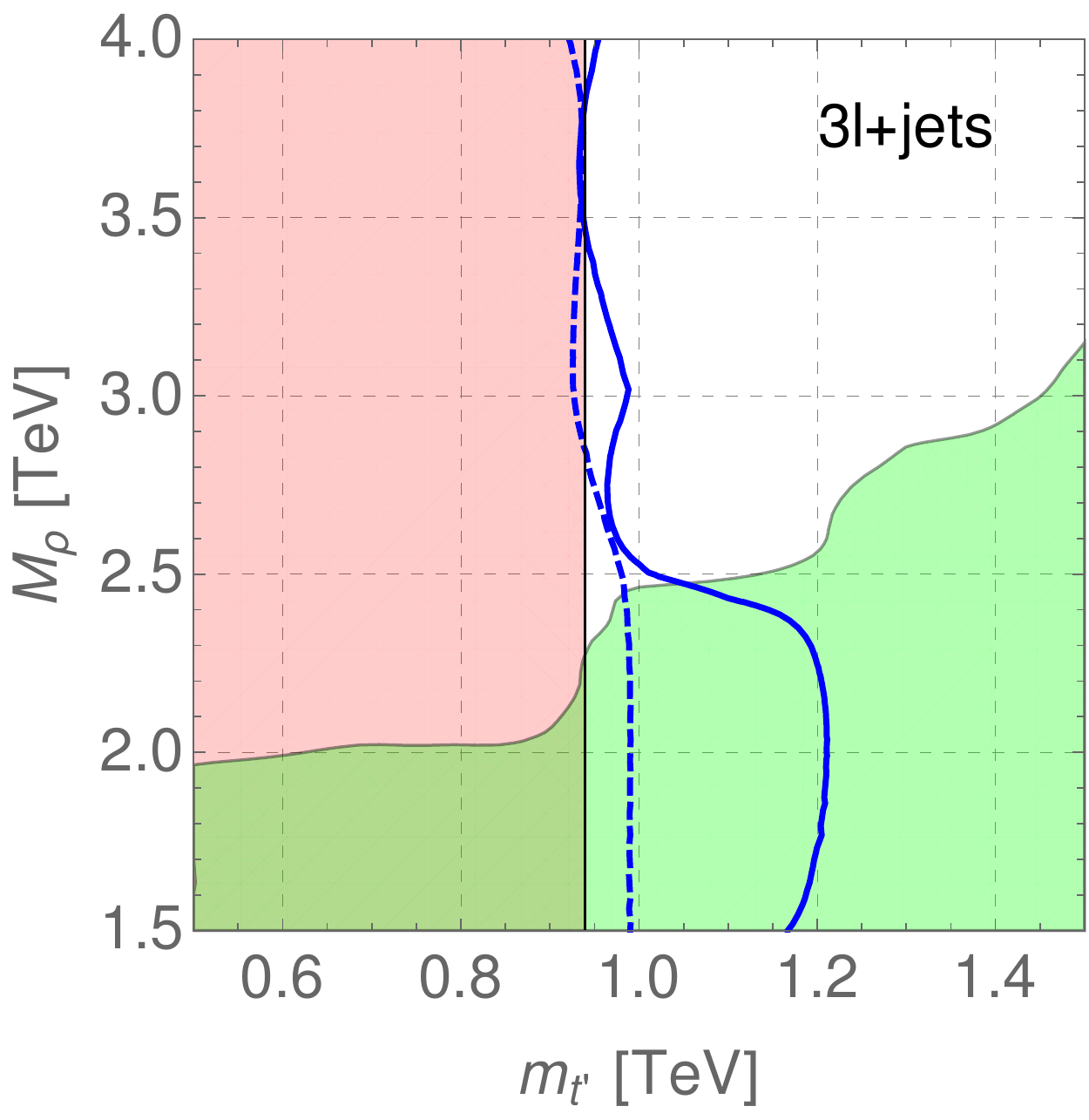}}
	\caption{(a) Feynman diagram and (b) $95\%$ C.L. exclusion contours from the trilepton channel on the ($m_t^{\prime},M_{\rho}$) plane. In (b) solid blue line is the plot for NWA, dashed blue line represents the plot for fat-width correction, the red excluded region corresponds to the QCD production of the top partner, and the green region is the $95\%$ C.L. exclusion region from KK-gluon search. The regions to the left of the contours are excluded.}
\end{figure*}

\subsubsection{Dilepton + $\geq$2J + jets}
The simulated process for the signal in our model is depicted in Figure~\ref{fig:process_feyn_di}. The dominant SM background subprocesses which contribute to the 2l + $\geq$2J + jets channel are $Z$+jets, $t\bar{t}$ and $t\bar{t}+X$ where $X$ is any massive gauge boson or the Higgs. The preselection cuts are exactly the same as those mentioned in the previous subsection. The relevant kinematic selections on the final state events for this analysis are as follows: 
\begin{itemize}
	\item All events were required to have exactly two oppositely charged same flavored leptons (electrons or muons) with $p_T$ greater than 28 GeV. Out of these the lepton pair decaying from the $Z$ boson was identified as the one with invariant mass closest to the $Z$ boson mass.
	\item The invariant mass of the same flavor charged lepton pair was required to be within 10 GeV of the $Z$ boson mass.
	\item The transverse momentum of the charged lepton pair ($p_{T_{l^+l^-}}$) was required to be greater than 250 GeV. This is to exploit the 	high $p_T$ feature of the $Z$ boson which comes from the decay of the heavy top partner.
	\item All events were required to have at least two small-R jets which were clustered using anti-$k_T$ algorithm with a radius parameter 0.4 and had $p_T$ greater than 25 GeV. At least two of them were required to be $b$-tagged which help in suppressing the SM background coming from $Z$+jets. 
	\item All events were required to have at least two large-R jets.
	\item A variable $H_T$ was defined as the scalar sum of the $p_T$ of all small-R jets and was required to be greater than 1150 GeV.
\end{itemize}
To validate the recast code of this search written in {\tt MadAnalysis5} we have generated SM $t\bar{t}$ and $t\bar{t}+X$ events and matched the cross section times efficiency times integrated luminosity with those given in Table 11 of Ref.~\cite{Aaboud:2018saj} within 10\% accuracy. The $95\%$ C.L. exclusion contour from this analysis is shown in Figure~\ref{fig:exclusion_di_2J}. As can be seen from the plot, the constraints on $m_{t^{\prime}}$ become more severe than the QCD limit (shaded red) for $M_{\rho}<3$ TeV. Part of this region is excluded from the direct limit on $M_{\rho}$ from the KK-gluon search (shaded green). Inclusion of the 1PI propagator significantly modulates the contribution of the $\rho$  on the bound on $m_{t^{\prime}}$ as can be seen by comparing the dashed (1PI propagator) and the solid (narrow-width) blue contours.

\subsubsection{Trilepton+jets}
In this section we summarize the constraints on the parameter space of our benchmark model from ATLAS 3l+jets study \cite{Aaboud:2018saj}. The simulated process for the signal in our model is depicted in Figure~\ref{fig:process_feyn_tri}. 
The dominant SM background subprocesses in this case are diboson production, $Z$+jets and $t\bar{t}$. 

Unlike the case of the dilepton channel, this final state is analyzed without a large R-jet and an additional charged lepton 
is present in the final state. Besides the preselection of an oppositely charged same flavor lepton pair coming from the $Z$ boson, an additional charged lepton is required in the final state. The cuts applied to mimic the ATLAS trilepton search \cite{Aaboud:2018saj} are described below.
\begin{itemize}
	\item All events were required to have at least three charged leptons (electron or muon) satisfying the preselection requirements on their minimum $p_T$ requirement of 28 GeV. Of these, a same flavor oppositely charged lepton pair was identified as the one coming from the decay of the $Z$ boson by requiring that the pair has an invariant mass closest to the $Z$ boson mass.
	\item As before, the invariant mass of the same flavor charged lepton pair was required to be within 10 GeV of the $Z$ boson mass.
	\item The $p_{T_{l^+l^-}}$  of the charged lepton pair satisfying the $Z$ boson mass window condition was required to be greater than 200 GeV.
	\item Events were required to have at least two jets clustered using anti-$k_T$ algorithm with a radius parameter 0.4 and having $p_T$ greater than 25 GeV. At least one of them was required to be $b$-tagged.
\end{itemize}
To validate the recast code of this search written in  {\tt MadAnalysis5} we have generated SM diboson and $Z$+jets events and matched the cross section times efficiency times integrated luminosity with those given in Table 13 (SR column) of Ref.~\cite{Aaboud:2018saj} within 10\% accuracy. The $95\%$ C.L. exclusion contour from this analysis is shown in Figure~\ref{fig:exclusion_tri}. As can be seen from the plot, the constraints on $m_{t^{\prime}}$ become more severe than the QCD limit (shaded red) for 
 $M_{\rho}< 2.5$ TeV, most of which is excluded from the direct limit on $M_{\rho}$ from the KK-gluon search (shaded green). The bound on $m_{t^{\prime}}$ shifts more towards the QCD limit when $\rho$ is considered fat.

\subsection{Future projection}
\begin{figure*}[t]
	\centering
	\subfloat[\label{fig:exclusion_total}]{\includegraphics[scale=0.5]{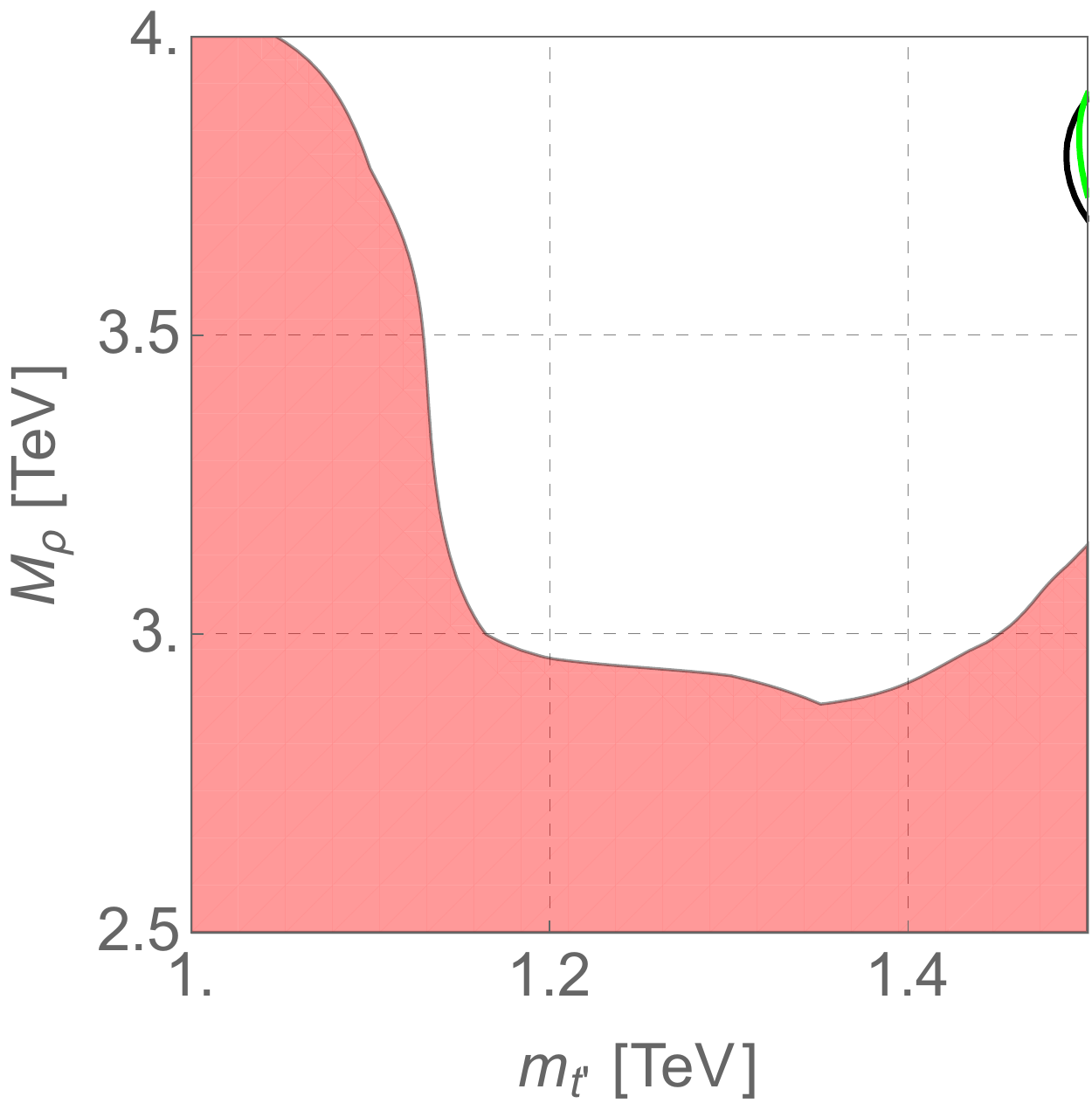}}\hspace{1cm}
	\subfloat[\label{fig:exclusion_total_interference}]{\includegraphics[scale=0.68]{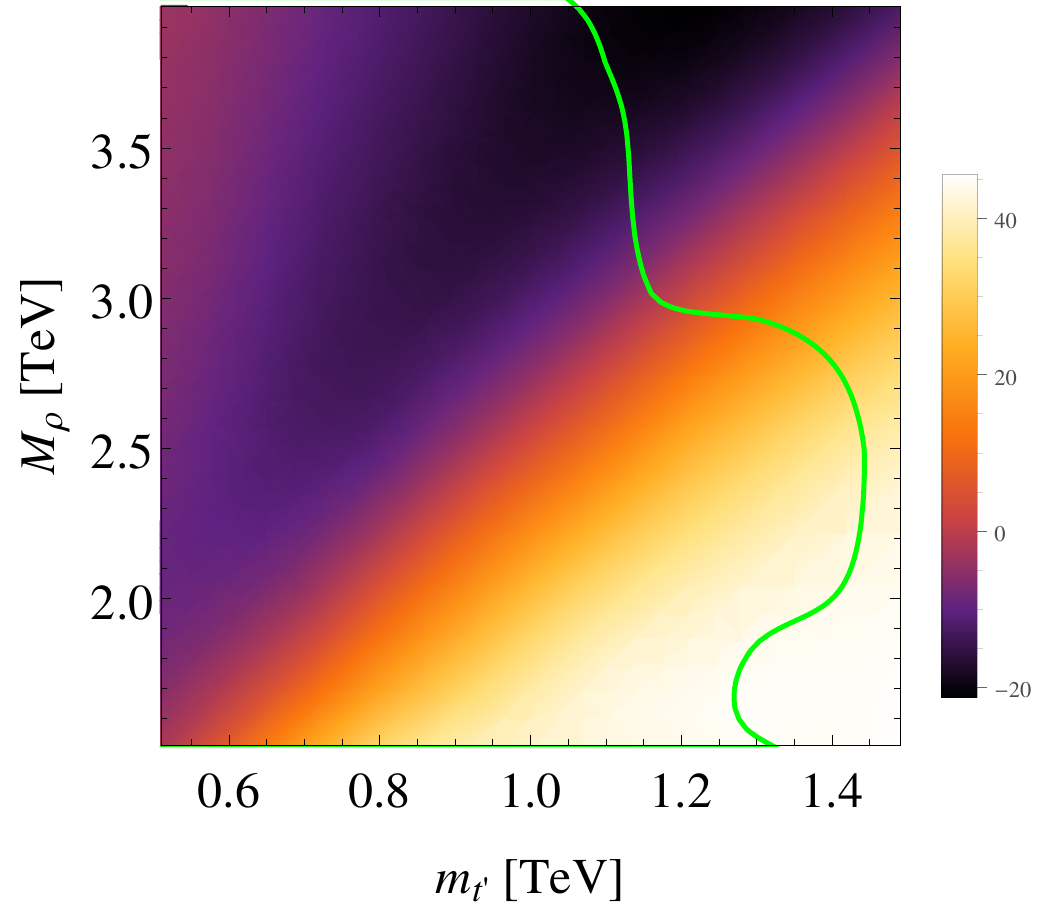}}
	\caption{(a) Summarized $95\%$ C.L. exclusion region from all three (mono,di,tri)lepton searches and from the KK-gluon search. Also plotted are the 300 $fb^{-1}$ projections from dilepton+1J (black) and dilepton+$>=$2J (green). (b) Density plot of the percentage contribution of the interference term in the top partner production cross section with present contour.}
\end{figure*}

LHC in its Run \uppercase\expandafter{\romannumeral3\relax} is expected to reach an integrated luminosity of 300 fb$^{-1}$ by the year 2023 before the third long shutdown. We  present the reach of these searches with the  projected integrated luminosity for 300 fb$^{-1}$.  A simplistic approach has been followed by scaling up the integrated luminosity keeping the cross section and efficiency unchanged. This is a very optimistic prediction as with increasing integrated luminosity, the increased pileup is expected to drop the efficiency which we do not take into account. A combination plot which shows the disallowed region from all the  present searches at 13 TeV described before and the 300 fb$^{-1}$ integrated luminosity projections of the dilepton searches are presented in Figure~\ref{fig:exclusion_total}. The projection of the trilepton search was subdominant and beyond the parameter space presented in Figure~\ref{fig:exclusion_total}. The projection of the monolepton search falls within the total current bound and thus we chose to omit it. The projections indicate significant enhancement of the bounds on the $M_{\rho} - m_{t^{\prime}}$ parameter space with increased integrated luminosity. We define interference $\mathcal{I}$ as 
\begin{equation}
\mathcal{I}=\frac{|\sigma-(\sigma_{\rho}+\sigma_{QCD})|}{\sigma}\times 100
\end{equation} 	
where $\sigma$ is the $pp\rightarrow t^{\prime}\bar{t^{\prime}}$ cross section. To demonstrate the impact of interference we present a density plot of $\mathcal{I}$ overlaid with the exclusion bounds in Figure~\ref{fig:exclusion_total_interference}. In the entire region of parameter space $\mathcal{I}$ can be as large as 40\%. 

\section{Conclusion}
In this paper we have revisited the constraints on the charge $2/3$, top-like,  top partner ($t^{\prime}$) from  the direct searches at LHC run 2 in the relatively clean  lepton(s) + jets final states. We study the impact on top partner searches from a massive colored vector boson resonance ($\rho_\mu$), the so-called gluon partner,  which is generic along with the top partners in a wide  class of models where electroweak symmetry breaking is driven by strong dynamics. We demonstrate how these constraints are modified if the $\rho_\mu$ is a broad resonance.  We recast the  monolepton+jets (CMS), dilepton+jets and trilepton+jets (ATLAS)  searches to put constraints on the parameter space of the model. Previous approaches to check the effect of fatness have rescaled the cross section keeping the signal efficiencies intact thus assuming that the final state kinematic shapes remain unaffected due to fatness. We have taken the effect of fatness in kinematic shapes into account by replacing the Breit-Wigner form of the $\rho_{\mu}$ propagator with its full 1PI resummed form in the event generation level. This also allows us to incorporate the impact of interference which is demonstrably significant in certain regions of parameter space. As can be seen from the resulting plots, the dilepton channel requiring at least two large-R jets provides the strongest bound on our parameter space and excludes $m_t^{\prime}$ up to 1.15 TeV for $M_{\rho}$=1.5 TeV when done using the proper 1PI propagator. The presence of $\rho$ with the proper 1PI propagator increases the bound on $m_{t^{\prime}}$ by up to 9\%. Significantly, implementation of the 1PI propagator  reduces the overestimated NWA bound by up to 24\% in certain regions of the parameter space. For values of $M_{\rho}$ greater than 2.5 TeV, the  contribution from the $\rho_\mu$ mediated process decouples  and the constraints essentially reduce to the limits obtained assuming pure QCD production of the top partners. 

\paragraph*{Acknowledgements\,:}
We thank Avik Banerjee and Benjamin Fuks for useful comments on our work. S.D. acknowledges MHRD, Govt. of India for the research fellowship. The work of T.S.R. is
partially supported by the Department of Science and Technology, Government of India,
under the Grant Agreement No. ECR/2018/002192  (Early Career Research Award). 
S.K.R.  acknowledges financial support from the Department of Atomic Energy, Government of India, for the
Regional Centre for Accelerator-based Particle Physics (RECAPP), Harish-Chandra Research Institute. T.S.R. would like to express a special thanks to the Mainz Institute for Theoretical Physics (MITP) of the Cluster of Excellence PRISMA+ (Project ID 39083149) for its hospitality and support.

\appendix

\section{Composite Higgs Effective Framework }
\label{appn:chframework}
While the Lagrangian for the top partner and gluon partner given in Equation \ref{master:lgn} is  phenomenological it can be embedded into the motivated composite Higgs framework. In this appendix we briefly sketch out the minimal framework that forms the basis of the simplified Lagrangian explored in this paper. The gauge hierarchy problem can be readily addressed by considering that the Higgs has a nontrivial extension in space. Such a composite object is naturally associated with a scale $f$ related to the size of the Higgs. However such extension results in serious modification of the Higgs coupling over the SM predictions. Essentially a composite Higgs with $f\sim v$ is ruled out by oblique electroweak parameters. This can however be circumvented if one assumes the Higgs as a pNGB of a strong sector. In the minimal realization such framework contains two distinct sectors with the usual elemental SM sector sans the Higgs on one hand and a strongly coupled sector where the dynamics results in spontaneous symmetry breaking of a continuous global symmetry that results in Nambu-Goldstone modes that can be identified with the Higgs doublet of the SM on the other.\\\\
A linear mixing between the operators of the strong sector and the SM states generates the Yukawa couplings for the Higgs states. This partial compositeness framework can be written as,
\begin{equation}
 {\mathcal{L}}_{mix} = \bar{\Psi}^i_L{\mathcal{O}}^i_{R} + \bar{\Psi}^i_R{\mathcal{O}}^i_{L} + h.c.
 \label{op:eq1}
 \end{equation}
 where $\Psi^i_{L/R}$ are the standard model fermions, $i$ is the flavor index and the ${\mathcal{O}}^i_{R}$ are operators of the strong sector that are in the $(3,1,2/3)$ representation of the SM gauge group. The operators are saturated by resonances of the strong sector, for example ${\mathcal{O}}^3_{L} \supset U_L + \ldots.$ The Lagrangian in Equation \ref{master:lgn} is obtained by assuming that the right-handed top mixes considerably with the strong sector resonances.
Such a minimal realization of the partial compositeness framework naturally necessitates  the existence of vector operators of the strong sector in the adjoint representation of the color $SU(3).$ We can define,
 \begin{equation}
  \mathcal{J}_{\mu} \equiv \bar{\mathcal{O}}\gamma_\mu \mathcal{O},
 \end{equation}
where $\mathcal{O}$ represents the fermionic operators defined in Equation~\ref{op:eq1}. And one can write down a linear mixing of the form
\begin{equation}
 \Delta {\mathcal{L}}_{mix} = \mathcal{J}_{\mu}G^\mu
\end{equation}
where $G^{\mu}$ are the SM gluons. However the large anomalous dimension of strong sector operator $ \Delta {\mathcal{L}}_{mix} $ makes them hopelessly irrelevant.  We will assume that the main coupling of the  gluon partner to the SM sector is through its couplings with the the top partners. The interactions in Equation \ref{master:lgn} are given by assuming $\mathcal{J}_{\mu} \supset \rho_\mu.$ The specific realization in Section \ref{sec:model} is obtained by assuming the top right is fully composite while the left chiral component of the top communicates to the strong sector through a linear coupling of the form given in Equation \ref{op:eq1}.
\bibliographystyle{h-physrev}
\bibliography{ref}

\end{document}